%% file: main.tex
\documentclass[%
reprint,
superscriptaddress,
nofootinbib,
 amsmath,amssymb,
prd,
floatfix,
]{revtex4-2}
\usepackage[dvipsnames]{xcolor}

\usepackage[utf8]{inputenc}
\usepackage{graphicx}
\usepackage{xcolor}
\usepackage{comment}
\usepackage{soul}
\usepackage{siunitx}
\usepackage{xspace}
\usepackage[mathlines]{lineno}

\newcommand{\eh}{$e^{-}h^{+}$\xspace}

\begin{document}

\title{Light Dark Matter Constraints from SuperCDMS HVeV Detectors Operated Underground with an Anticoincidence Event Selection}

\input{SuperCDMS_AuthorList}

\date{\today}

\begin{abstract}

This article presents constraints on dark-matter-electron interactions obtained from the first underground data-taking campaign with multiple SuperCDMS HVeV detectors operated in the same housing. An exposure of 7.63~g-days is used to set upper limits on the dark-matter-electron scattering cross section for dark matter masses between 0.5 and 1000\,MeV/$c^2$, as well as upper limits on dark photon kinetic mixing and axion-like particle axioelectric coupling for masses between 1.2 and 23.3\,eV/$c^2$. Compared to an earlier HVeV search, sensitivity was improved as a result of an increased overburden of 225~meters of water equivalent, an anticoincidence event selection, and better pile-up rejection. In the case of dark-matter-electron scattering via a heavy mediator, an improvement by up to a factor of 25 in cross-section sensitivity was achieved.

\end{abstract}

\maketitle


\section{Introduction}

    
    As the long-standing weakly interacting massive particle (WIMP) paradigm of dark matter (DM) has become increasingly constrained by experimental results, well motivated sub-GeV DM models have gained more focus \cite{Essig:2016}. Such light DM models introduce a new gauge boson that either serves as a force mediator between thermal DM and Standard Model particles or comprises the relic DM abundance on its own. Experimentally, the former can be detected via DM-electron scattering and the latter via electron absorption processes \cite{Battaglieri:2017}.
 
    SuperCDMS high-voltage eV-resolution (HVeV) detectors use the Neganov-Trofimov-Luke (NTL) effect \cite{Neganov:1985, Luke:1988} to detect electron interaction energies as low as the band gap energy of silicon (Si). In the previous above-ground DM searches with HVeV detectors, constraints on the DM-electron scattering cross section, the dark photon kinetic mixing, and the axioelectric coupling of axion-like particles (ALPs) were obtained \cite{Agnese:2018, Amaral:2020ryn}. The limiting factor for the DM sensitivity in these searches was the excess of background events, presumably caused by luminescence of materials in the detector holder \cite{0vev}.
    
    In this work, we analyze data from the first underground HVeV DM search, where the underground location reduced the rate of cosmogenic background events. To further reduce background from external sources, we operated multiple HVeV detectors in the same housing and applied an anticoincidence event selection. These changes allowed us to improve upon the DM-electron interaction constraints obtained in the previous HVeV experiments.

\section{Experimental Setup}\label{sec:setup}

    The data for this analysis were acquired at the Northwestern EXperimental Underground Site (NEXUS) at Fermilab during January and February 2021 (also referred to as HVeV Run 3). The NEXUS facility provides 225~meters of water equivalent rock overburden \cite{minos_2015}. Detectors were operated in a dilution refrigerator with dedicated lead shielding to reduce the radioactive background from the cavern where NEXUS is located. A light-tight copper housing containing the detector payload was thermally coupled to the mixing chamber of the cryostat. The temperature of the mixing chamber was stabilized at 10.5\,mK throughout the run.

    The detector payload consisted of four SuperCDMS HVeV detectors: one identical to the detector used in Ref.~\cite{Amaral:2020ryn} and described in detail in Ref.~\cite{ren:2020} (labeled NF-C), two similar detectors with different QET\footnote{Quasiparticle-trap-assisted Electrothermal-feedback Transition-Edge Sensor \cite{QETs_1995}} configurations (labeled NF-H and NF-F), and one identical to the detector used in Ref.~\cite{Agnese:2018} (labeled R1). Data from the NF-F detector were not used since a flaw in its sensor layout was discovered that prevented it from being operated properly. Each detector was fabricated on a 0.93~g high-purity Si wafer ($10 \times 10 \times 4$~\SI{}{mm^3}). Two concentric square phonon-sensing QET channels of equal areas covered one side of each detector. A voltage differential of $V_{\text{bias}}$ = \SI{100}{V} was applied between each detector's QET-bearing surface and the opposing surface to induce NTL amplification.

\begin{figure}[!htb]
        \centering
        \includegraphics[width=1\linewidth]{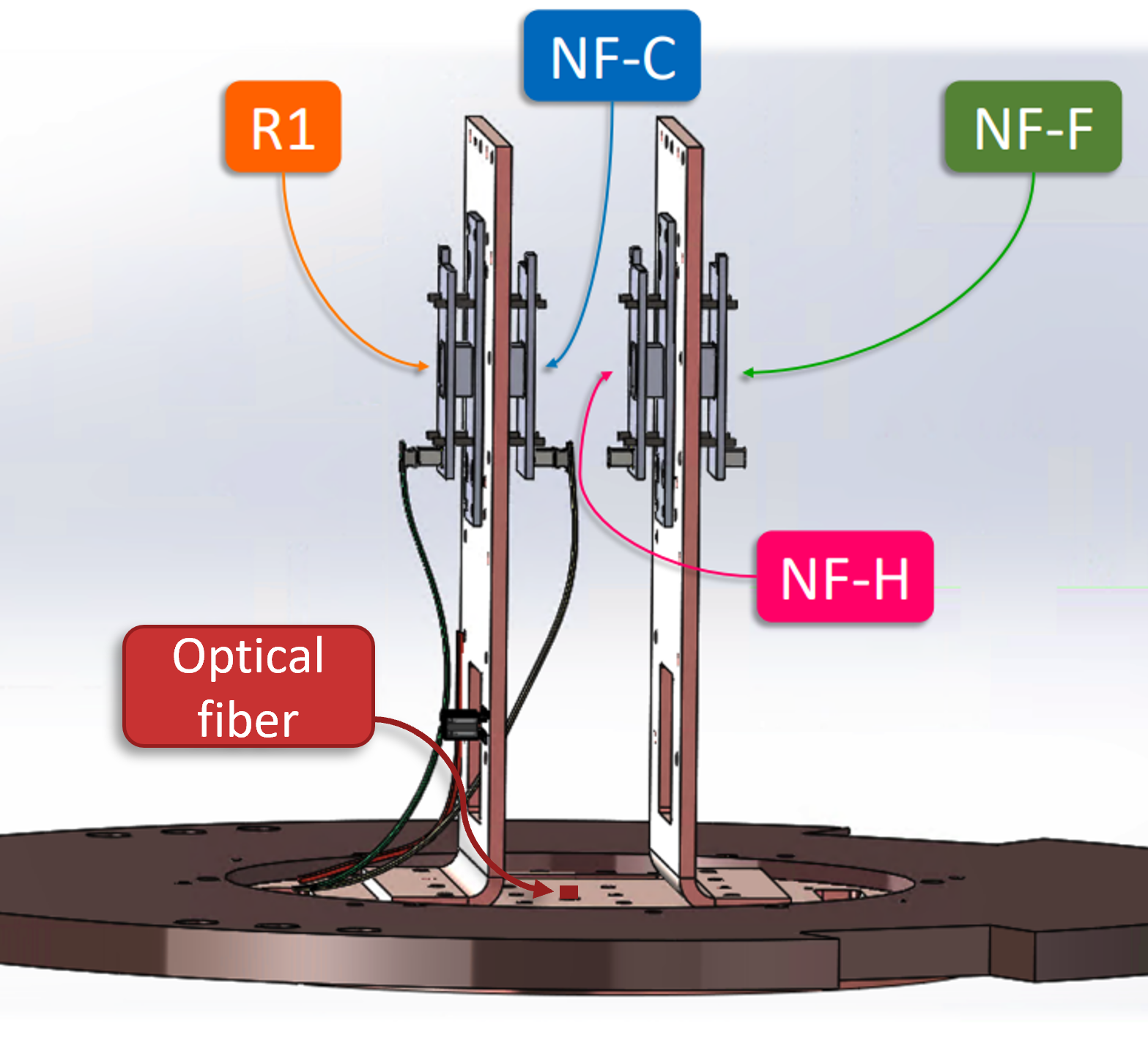}
        \caption{Illustration of the detector setup. Four HVeV detectors were mounted on two copper holders that were thermally anchored to the mixing chamber of a dilution refrigerator. Each detector was clamped between two printed circuit boards. An optical fiber coupled to the laser was inserted between the detector holders. Individual MDM25 connectors were used with each detector. The associated readout cables are shown for R1 and NF-C.}
        \label{fig:payload}
    \end{figure}

    As shown in Fig.~\ref{fig:payload}, the detectors were mounted in pairs on two copper holders, with R1 and NF-C on one holder, and NF-F and NF-H on the other. Each detector was clamped between two printed circuit boards (PCBs). Only NF-C was used directly for the DM search due to its superior performance and because even with a single detector we expected the sensitivity to be limited by background rather than exposure, despite the rock overburden and anticoincidence selection. NF-H and R1 served as anticoincidence detectors, as described in Section~\ref{sec:selection}.

    The detectors were read out using a SuperCDMS Detector Control and Readout Card (DCRC) with \SI{625}{kHz} sampling frequency. Data were acquired in 0.5-second-long intervals without online triggering. Due to DCRC limitations on data throughput, these intervals were separated by dead-time intervals of random lengths between 0 and 1~second, resulting in an overall dead time of 50\%. An offline trigger was applied later, as described in Section~\ref{sec:reconstruction}. In total, 13.14~days of DM-search data were collected during the run. The full data sample was used to define the event selections described in Section~\ref{sec:selection}, but only 10\% of the sample was used to define specific parameters that could have introduced subjective bias into the analysis described in Section~\ref{sec:results}.

    As in the previous HVeV operations, a \SI{1.95}{eV} laser was used as a source of calibration events. Photons from the laser were transmitted into the detector housing through an infrared filter and an optical fiber inserted into the bottom of the housing between the detector holders. Several calibration datasets with varied laser intensity were acquired throughout the run.

\section{Event Reconstruction and Calibration}\label{sec:reconstruction}

    To identify and characterize energy depositions, we summed together the QET currents in a given detector's inner and outer channels after correcting for differences in the channels' responsivities. We then applied a Gaussian derivative filter, defined as a convolution with the first derivative of a Gaussian-shaped signal. A trigger threshold corresponding to $\sim$25\% of the energy produced by a single \eh-pair was then applied to the filtered signals in NF-C and NF-H. In the case of R1, the threshold was set to $\sim$50\% of the energy produced by a single \eh-pair due to the higher noise level observed in this detector. Trigger times were corrected using a constant time offset to approximately correspond to the rising edge of the unfiltered pulse. Time windows spanning $\pm$\SI{6.55}{ms} around each corrected trigger point were further processed to identify the amplitude of each pulse.

    Pulse amplitudes and frequency-domain goodness of fit $\chi^2$ values were calculated with the Optimal Filter (OF) algorithm \cite{Wilson:2022}. The OF pulse template for each detector was constructed by averaging pulses from the laser calibration data. The noise power spectral density (PSD) used in the OF to optimally weight the signal frequency components was calculated separately for each minute of data to mitigate the effect of a changing noise environment. When solving the OF $\chi^2$ minimization problem, the template was allowed to move around the trigger point within $\pm$15 time samples ($\pm$\SI{24}{\micro s}) to eliminate any remaining trigger misalignment.
    
    In addition to the standard OF template fit, we performed a pile-up OF fit to each triggered event \cite{Kurinsky:2018}. This OF variant simultaneously fits the sum of two otherwise identical templates with independent amplitudes and variable time shifts. One time shift was limited to be within $\pm250$ time samples ($\pm$\SI{400}{\micro s}) of the trigger point. The other time shift was limited to be within $\pm250$ time samples of the first. The $\chi^2$ values of these fits were used to identify pile-up events as described in the next section.

    As in Ref.~\cite{Amaral:2020ryn}, we calibrated OF fit amplitudes to event energies using the alignment of laser-data \eh-pair peaks in the QET current distributions with their expected energies:

    \begin{equation} \label{eq:calib_peak_energy}
        E_{n} = n ( 1.95 \text{\,eV} + e V_{\text{bias}} ),
    \end{equation}
    where $n$ is the number of photons hitting the detector each producing one \eh pair, \SI{1.95}{eV} is the photon energy, $e$ is the absolute electron charge, and $V_\text{bias}$ is the bias voltage. We set the energy region of interest (ROI) for this analysis to a range that covers the first four \eh-pair peaks, 50--440\,eV, as this range was found to be the most sensitive to sub-GeV DM-electron recoil interactions. 

\section{Data Selection}\label{sec:selection}

\begin{figure}[!htb]
        \centering
        \includegraphics[width=1\linewidth]{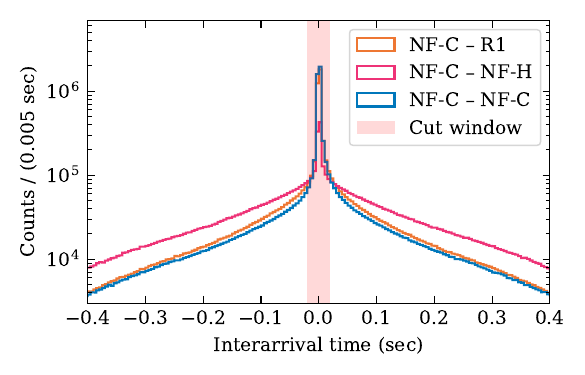}
        \caption{Distributions of time intervals from each triggered event in NF-C (reference event) to the closest-in-time events in R1 (orange), NF-H (pink) or NF-C (blue). The values are positive (negative) when the reference event is after (before) the closest event in the respective detector. Strong time correlations are observed as sharp peaks around zero seconds. The red-shaded window shows the $\pm$20\,ms window excluded by the anticoincidence live-time selection.}
        \label{fig:dt}
    \end{figure}

    DM events are expected to be random and independent of each other, therefore the times between DM events occurring in a single detector are expected to follow an exponential distribution. Moreover, due to the low expected interaction cross sections, the probability of the same DM particle interacting in multiple detectors is negligible (or zero for the absorption models). Therefore, the time intervals between DM events in different detectors are also expected to follow an exponential distribution. However, as shown in Fig.~\ref{fig:dt}, the distribution of interarrival times of events occurring in NF-C, as well as the distributions of time intervals between events in NF-C and the closest-in-time events in the other two detectors, contain sharp peaks around 0 seconds indicating events correlated in time that are inconsistent with DM candidates. We remove such correlated events by rejecting events in any detector that fall within 20-ms-long windows before and after each event. This ensures that events less than \SI{20}{ms} apart are excluded from the analysis and the respective live time is removed, regardless of whether the events occurred in the same or different detectors. The choice of \SI{20}{ms} was made to remove the majority of correlated events, while minimizing the loss of live time. This anticoincidence selection removed 22.9\% of the total live time, while reducing the rate of events creating more than one \eh pair by an order of magnitude. 

    In addition to the anticoincidence live-time selection, live-time selections based on the stability of temperature and the QET current baseline were applied to ensure energy amplification stability. The remaining live time after applying all the selection criteria amounted to 9.12~days. Taking into account the mass of the NF-C detector (0.93\,g) and the fact that only 90\% of the data are used for the limit setting, the DM-search exposure was \mbox{7.63~g-days}.
    
    To remove poorly reconstructed events, we applied two selection criteria based on the OF fit quality. One selection criterion was based on the $\chi^2$ value of the one-pulse OF fit, while the other was based on the difference between the $\chi^2$ values of the one-pulse and two-pulse OF fits. The former criterion rejects events with pulses that have shapes deviating from the pulse template, while the latter is especially effective at rejecting events with overlapping pulses. Both criteria were developed using the laser calibration data in a way that ensures an energy-independent selection efficiency of 95\%. When both criteria were applied, the selection efficiency was measured to be 90\% for all events in the energy ROI.
    
\begin{figure}[!htb]
        \centering
        \includegraphics[width=1\linewidth]{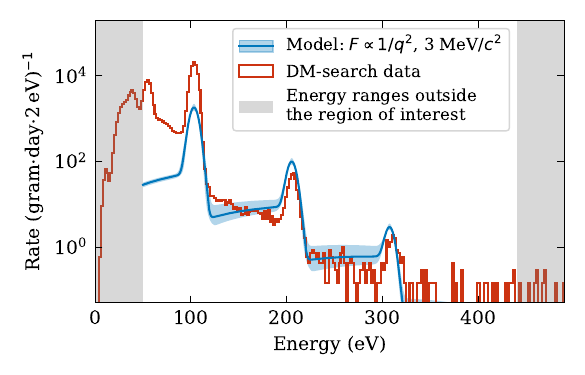}
        \caption{The red curve shows the event rate for the dark-matter search data after live-time and event-based data selection, corrected by the selection efficiency. The blue curve shows an example of a dark-matter-electron recoil signal model for the case of \SI{3}{MeV c^{-2}} dark matter interacting via an ultra-light mediator (form-factor $F\propto1/q^2$). The signal model is shown for the cross section value corresponding to the 90\% confidence level limit produced in this work. The blue-shaded area shows the systematic uncertainties on the signal model dominated by the impact ionization uncertainty (see Section~\ref{sec:signal_models}). The gray-shaded regions mark the energy ranges outside the region of interest.
        }
        \label{fig:spectrum}
    \end{figure}

    Figure~\ref{fig:spectrum} shows the DM-search-data energy spectrum with the live-time and event-based selections applied, corrected by the selection efficiency. Three \eh-pair peaks are visible in the spectrum, with several more peaks below the first \eh-pair peak, which is located at $\sim$100\,eV. One hypothesized explanation for the sub-\eh peaks is charge trapping on the sidewalls of the detector; however, further studies are required to better understand the exact origin of these events.
    
\section{Dark matter signal models}\label{sec:signal_models}

    We used the resulting energy spectrum to set exclusion limits on various DM model parameters. We used the DM-electron scattering model described in Ref.~\cite{Essig:2016} to set limits on the effective DM-electron cross section ($\bar{\sigma}_e$), and we used the dark-photon and ALP models described in Ref.~\cite{Hochberg:2017} to set limits on the kinetic mixing parameter ($\varepsilon$) and axioelectric coupling constant ($g_{ae}$).

    For each model, we assumed a local DM density of \SI{0.3}{GeV c^{-2} cm^{-3}} that exclusively consisted of the candidate particle. For DM-electron scattering, we assumed the DM velocity distribution described in Ref.~\cite{Baxter2021}. Specifically, we used the average DM velocity in the galactic frame of 238.0\,km/s and the galactic escape velocity of 544.0\,km/s. We also calculated the average Earth velocity (with respect to the DM halo) of 253.7\,km/s using the average DM velocity in the galactic frame, the solar peculiar velocity, and the year-averaged velocity of Earth with respect to the Sun from Ref.~\cite{Baxter2021}.

    The dark-photon and ALP rates from Ref.~\cite{Hochberg:2017} are proportional to the real part of the complex conductivity ($\sigma_1$) of the target material. As was done for Ref.~\cite{Amaral:2020ryn}, we write $\sigma_1$ in terms of the photoelectric absorption cross section ($\sigma_\mathrm{pe}$), the real index of refraction of Si ($n_{\text{Si}}$), and the density of Si ($\rho_{\text{Si}}$):
    
    \begin{equation} \label{eq:complex_conductivity}
        \sigma_1 = n_{\text{Si}} \sigma_\mathrm{pe} \rho_{\text{Si}}.
    \end{equation}
    
    For this analysis, we used $\sigma_\mathrm{pe}$ from Refs.~\cite{Stanford2021,Edwards:1997,Henke:1993,XCOM} with the specific reference dependent on the absorption energy. We used the Si band gap energy ($E_\mathrm{gap}$) of \SI{1.131}{eV}, calculated by performing a revised fitting of the data in Ref.~\cite{Stanford2021} in order to properly account for the uncertainties in the $\sigma_\mathrm{pe}$ model (further details can be found in Appendix~\ref{Egap_app}).
    
    For the Si index of refraction $n_{\text{Si}}$, we used the energy dependent values from Ref.~\cite{Edwards:1997}. Note that in the previous HVeV analysis~\cite{Amaral:2020ryn}, we used Ref.~\cite{Pospelov:2008a} for the ALP absorption model in which $n_{\text{Si}}=1$ is assumed. To correct for this difference when comparing the limits on the axioelectric coupling $g_{ae}$, we divided the limit from Ref.~\cite{Amaral:2020ryn} by the square root of $n_{\text{Si}}$, as for a fixed ALP absorption rate, $g_{ae}^2$ is inversely proportional to $\sigma_1$ which is in turn proportional to $n_{\text{Si}}$~\cite{Hochberg:2017}.

    The total phonon energy measured for an event ($E_{ph}$) is the recoil or absorption energy ($E_r$) plus the energy produced by the NTL effect:

    \begin{equation} \label{eq:phonon_energy}
        E_\mathrm{ph} = E_\mathrm{r} + n_{eh} e V_{\text{bias}}
    \end{equation}

    \noindent where $n_{eh}$ is the number of \eh pairs generated by the event. For this analysis, we used the \eh-pair generation probabilities derived in Ref.~\cite{Ramanathan2020} to compute the distribution for $n_{eh}$ as a function of $E_\mathrm{r}$. For $E_\mathrm{r}$ up to 50\,eV, we interpolated the values provided in the supplementary materials of Ref.~\cite{Ramanathan2020} to estimate the probabilities for $E_\text{gap}$ equal to \SI{1.131}{eV}. For $E_\mathrm{r}$ above 50\,eV, we used Eqs.~13, 14, and 15 from Ref.~\cite{Ramanathan2020} with a small modification as described in Appendix~\ref{ion_app}.
    We included the following detector response effects into the signal models: bulk charge trapping (CT), bulk impact ionization (II), and energy resolution. For the bulk CT and II effects, we used the model described in Ref.~\cite{new_ctii}. The probabilities of CT and II processes to occur to a charge traversing the full thickness of the detector were obtained by fitting the model to the calibration data from Ref.~\cite{Amaral:2020ryn} and were found to be $(12.8\pm1.5)\%$ and $(1.6^{+1.8}_{-1.6})\%$, respectively.
    
    The energy resolution was modeled as an energy-independent Gaussian smearing, with a systematic uncertainty assigned to the nominal resolution value to account for possible energy dependency. For the nominal resolution value (4.26\,eV) we used the average between the low and high resolution estimates. The low resolution estimate (3.03\,eV) was defined by the baseline resolution found by fitting a Gaussian to the energy distribution of randomly triggered noise events. For the high resolution estimate (5.49\,eV) we used the Gaussian widths of the widest \eh-peak in the laser data within the energy ROI. The difference between the low and high estimates was used as the systematic uncertainty.
    
    An example of a DM-electron recoil signal model with the detector response effects applied is shown in Fig.~\ref{fig:spectrum}. The signal model shown is for a DM particle with a mass of 3\,MeV/$c^2$ scattering off of electrons via an ultra-light mediator (form-factor $F\propto1/q^2$). The systematic uncertainties considered are the uncertainties on the CT and II probabilities, resolution, selection efficiency and calibration. In the case of DM absorption models with particle energies $\lesssim \SI{3.2}{eV}$, the uncertainty on $\sigma_\mathrm{pe}$ (from measurements described in Ref.~\cite{Stanford2021} and the revised fit described in Appendix~\ref{Egap_app}) is also taken into account.
    
\section{Results}\label{sec:results}

    We applied the same limit-setting method as in Ref.~\cite{Amaral:2020ryn}. First, for each DM model and mass, we used 10\% of the experimental data to estimate the upper limits on interaction strength attainable using the remaining 90\% of the data. Limits were calculated by applying the Poisson counting method to events in each energy region enclosing an individual \eh-pair peak (defined using the signal models). Next, for each DM model and mass, we selected the \eh-pair peak expected to produce the most stringent limit. If more than one peak produced comparably strong limits, we allowed more than one \eh-pair peak to be selected. If only one peak was selected, the final interaction strength upper limit was calculated as the 90\% confidence level limit using the remaining 90\% of the data. If more than one peak was selected, the limits in each were calculated using the remaining 90\% of the data with the confidence level 
    \begin{equation}
        C = 0.9^{1/n},
    \end{equation}
    where $n$ is the number of selected peaks. This confidence level correction allowed us to choose the most stringent limit among the limits from the individual \eh-pair peaks, producing the final 90\% confidence level limit.
    
    Systematic uncertainties were not taken into account in calculating the confidence level limits. To estimate the effect of systematic uncertainties, we varied the signal model and experimental parameters within their systematic uncertainties, calculated limits for each variation using the method described above, and produced a band that includes 68.2\% of the resulting limit variations. For all models and masses, the resulting band boundaries are within 25\% of the nominal limit.

    \begin{figure*}[htbp]
        \centering
        \includegraphics[width=0.95\textwidth]{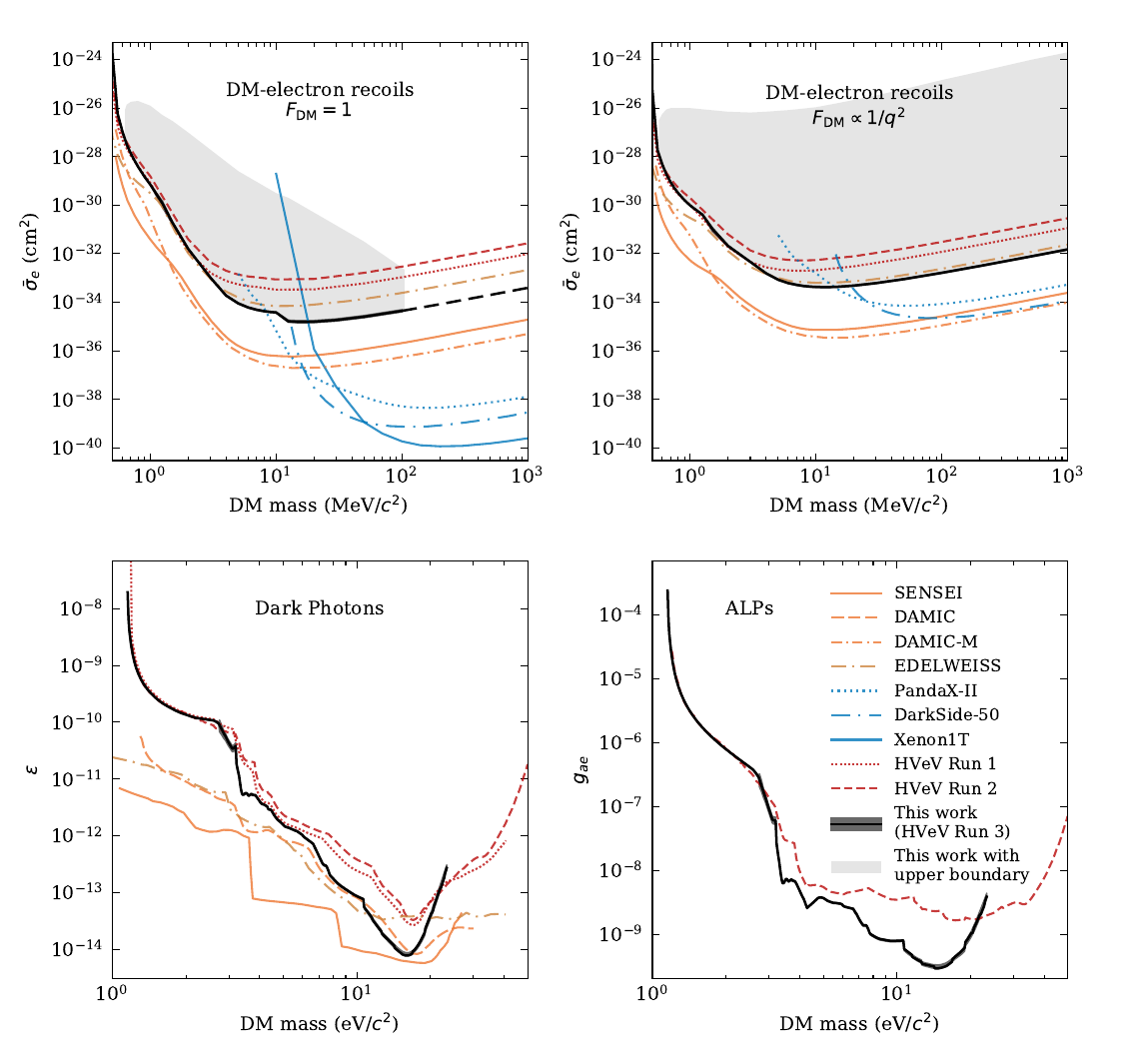}
        \caption{90\% confidence level limits on the DM-electron recoil cross section with form-factors $F_\mathrm{DM}=1$ (upper left) and $F_\mathrm{DM}\propto1/q^2$ (upper right), dark photon kinetic mixing parameter (lower left) and axion-like particle (ALP) axioelectric coupling constant (lower right). Limits produced in this work are shown in black, with dark gray bands around them showing the sensitivity of the limits to systematic uncertainties (though the band is too narrow to see for most of the models and masses). In the upper plots, light gray shaded areas show the exclusion regions estimated in this work with the upper boundary arising from the Earth shielding effect. Above $\sim$100\,MeV/$c^2$, the $F_\mathrm{DM}=1$ limit is shown as a dashed line to indicate the predicted loss of sensitivity due to the Earth shielding effect. Other limits, shown as colored lines, are from SENSEI \cite{Barak:2020}, DAMIC \cite{Aguilar-Arevalo:2019}, DAMIC-M \cite{damic_m}, EDELWEISS \cite{Arnaud:2020}, PandaX-II \cite{panda21}, DarkSide-50 \cite{darkside_er}, XENON1T \cite{Aprile:2019}, SuperCDMS HVeV Run~1 \cite{Agnese:2018}, and HVeV Run~2 \cite{Amaral:2020ryn}. The HVeV Run~2 ALP limit is scaled as discussed in Section~\ref{sec:signal_models}.}
        \label{fig:limits}
    \end{figure*}

    Figure~\ref{fig:limits} shows the resulting limits on the four considered DM models' signal strengths compared to the existing limits from other direct search experiments. The step in the DM-electron recoil limit with the form factor $F_\mathrm{DM}=1$ at the mass of 10\,MeV/$c^2$ is caused by the transition from the third to the fourth \eh peak in the limit-setting procedure. The dark photon and ALP limits do not extend above $\sim20$\,eV/$c^2$ due to a more limited energy ROI compared to the previous HVeV DM search~\cite{Amaral:2020ryn}. 

    Interactions of DM particles with the Earth's atmosphere and crust attenuate the expected DM flux and change the DM velocity distribution, limiting the experiment's capacity to exclude high interaction strength values. To estimate the upper boundary of the exclusion region of the DM-electron scattering cross section, we used a Monte Carlo simulation of the Earth crust nuclear stopping power described in detail in Ref.~\cite{damascus2019} and implemented in Ref.~\cite{DaMaSCUScrust}. A number of simplifications were made in this estimation: the DM flux was assumed to be directed vertically from above reaching the detector via the shortest path through the overburden, the standard continental crust composition was used (the default option in Ref.~\cite{DaMaSCUScrust}), the atmosphere stopping power was ignored as subdominant~\cite{damascus2019}, and lastly, the effects of CT, II and detector resolution were ignored. Moreover, the limit-setting technique implemented in Ref.~\cite{DaMaSCUScrust} is different from the one used to calculate the lower boundaries of the exclusion regions in this analysis. The Ref.~\cite{DaMaSCUScrust} limit-setting energy regions are one-sided intervals above a specified threshold rather than two-sided intervals around each \eh-pair peak. An older version of the Si crystal form-factor is used in Ref.~\cite{DaMaSCUScrust}. Additionally, no confidence level correction is included when combining Ref.~\cite{DaMaSCUScrust} limits from different energy regions. Therefore, the calculated upper boundary of the exclusion region, shown in Fig.~\ref{fig:limits} as a light gray shaded area, is an approximate estimate of the Earth shielding effect ignoring various higher order effects.
    
    One notable result of the Earth shielding effect is the predicted loss of sensitivity of our experiment to DM masses above $\sim$100\,MeV/$c^2$ for the heavy mediator case ($F_\mathrm{DM}=1$).
    Because uncertainties in the Earth shielding model have not been taken into account, neither the upper boundary of the exclusion region nor the location of the sensitivity loss are conservatively estimated.
    
    For the dark photon and ALP absorption scenarios, the upper boundaries of the parameter space exclusion regions were estimated to be above $10^{-7}$ and $10^{-2}$, respectively. A more accurate calculation would require knowledge of the composition of Earth's crust and atmosphere above the NEXUS facility, including various materials' complex conductivities that are unknown for energies corresponding to the DM masses of interest.

\section{Discussion and Outlook}

    In the analysis presented here, the anticoincidence event selection combined with the rock overburden and improved pile-up rejection allowed us to significantly improve the exclusion limits on the DM-electron scattering cross section, dark photon kinetic mixing and ALP axioelectric coupling compared to the previous HVeV results. The largest improvement, up to a factor of 25, is obtained for the DM-electron scattering via a heavy mediator for DM masses above 10\,MeV/$c^2$.

    The large number of background events that arrive closely spaced in time in different detectors suggests that the dominant source of background in this experiment is external, as opposed to detector internal sources such as stress-induced crystal microfractures reported in Ref.~\cite{anthony_2022}. This observation is in agreement with the hypothesis introduced in Ref.~\cite{0vev} that suggests luminescence from PCBs used in the detector holders to be a possible dominant source of background in the previous HVeV experiments. For the following HVeV run, we designed a detector holder with minimal luminescent material in order to reduce the hypothesized background. Analysis of the data collected in the run with the new holder is an ongoing effort.
    
\begin{acknowledgements}
We would like to thank Tongyan Lin and Tien-Tien Yu for valuable discussions of the DM signal models, Karthik Ramanathan for the help with the ionization model, and Timon Emken for the help with configuring the Earth shielding effect simulation.

We gratefully acknowledge funding and support received from the National Science Foundation, the U.S.\ Department of Energy (DOE), Fermilab URA Visiting Scholar Grant No.\~15-S-33, Natural Sciences and Engineering Research Council of Canada (NSERC), the Canada First Excellence Research  Fund, the Canada Foundation for Innovation, the Arthur B.~McDonald Institute (Canada), the Department of Atomic Energy Government of India (DAE), the Department of Science and Technology (DST, India), Deutsche Forschungsgemeinschaft (DFG, German Research Foundation) under Project No.\~420484612 and under Germany’s Excellence Strategy - EXC 2121 ``Quantum Universe''. Femilab is operated by Fermi Research Alliance, LLC, SLAC is operated by Stanford University, and PNNL is operated by the Battelle Memorial Institute for the U.S.\ Department of Energy under contracts DE-AC02-37407CH11359, DE-AC02-76SF00515, and DE-AC05-76RL01830, respectively. This research was enabled in part by support provided by Compute Ontario (\url{computeontario.ca}) and the Digital Research Alliance of Canada (\url{alliancecan.ca}). We acknowledge the funding support provided by the Mitchell Institute at Texas A\&M University to fabricate these detectors.
\end{acknowledgements}

\appendix

\section{Revised Photoelectric Model Parameters}\label{Egap_app}
In Ref.~\cite{Stanford2021}, a temperature-dependent model of indirect photon absorption in Si was fit to newly acquired measurements of the photoelectric cross section. These measurements were made for photons with energies between 1.21 and 2.77\,eV, and at four distinct temperatures between 0.5 and 295\,K. An indirect absorption model with four parameters was fit to the data. The parameters were: the band gap energy of the first indirect band gap $E_{g1}(0)$, the proportionality constant corresponding to the second indirect band gap $A_{2}$, and the coefficients $c_{0}$ and $c_{1}$ that parameterize the temperature-dependent proportionality constant of the first indirect band gap $A_{1}(T)$. The temperature dependence on $A_{1}(T)$ is described by: 
\begin{equation}\label{eq:Egap_app_A1}
A_{1}(T) = c_{0}e^{-c_{1}T}.
\end{equation}

The fit procedure performed in Ref.~\cite{Stanford2021} obtained the fit parameters in two steps. First, the data were fit to the model by allowing $E_{g1}(0)$, $A_{2}$, and four $A_{1}$ values (corresponding to the measurements taken at each temperature) to float. Next, the fit was performed again by keeping $E_{g1}(0)$ and $A_{2}$ fixed, and replacing $A_{1}$ with the right side of Eq.~\ref{eq:Egap_app_A1} in order to determine $c_{0}$ and $c_{1}$.

The analysis performed in this work investigates the effect of systematic uncertainties on the limit result, which includes the uncertainty of the photoelectric cross section due to the indirect absorption model. This requires proper propagation of the uncertainty in the indirect absorption model using the covariance matrix of the fit parameters. However, the covariance matrix for the fit was not provided in Ref.~\cite{Stanford2021}. Thus, a revised fit to the data in Ref.~\cite{Stanford2021} was performed for this work to obtain the covariance matrix of the fit parameters. Here, the indirect absorption model is fit to the data in Ref.~\cite{Stanford2021}, where $A_{1}(T)$ is again described by Eq.~\ref{eq:Egap_app_A1}, and the best-fit values of $E_{g1}(0)$, $A_{2}$, $c_{0}$ and $c_{1}$ are obtained in one step. The results of the fit are: $E_{g1}(0) = 1.131(3)$\,eV, $A_{2} = 7(2) \times 10^{3}$\,eV$^{-2}$cm$^{-1}$, $c_{0} = 321(7)$\,eV$^{-2}$cm$^{-1}$, and $c_{1} = 1.6(1) \times 10^{-3}$\,K$^{-1}$.

\pagebreak

The best-fit values obtained here were used in the analysis presented in this work to compute the indirect absorption model in Si. In order to propagate the uncertainty from this model, we used the variance and covariance values obtained from the revised fit and displayed in Table \ref{tab:Egap}.


\begin{table}[hbt!]
\centering
\caption[label]{Variance and covariance parameters from the revised, single-stage fit of the data and indirect absorption model described in Ref.~\cite{Stanford2021} to Eq.~\ref{eq:Egap_app_A1}.}
\begin{tabular}{llll}
\hline
Variance \hspace{0.7cm} & & \\
\hline
$\sigma^{2}_{c_{0}}$ & 42.6 & eV$^{-4}$cm$^{-2}$\\
$\sigma^{2}_{c_{1}}$ & \SI{9.38e-9}{} & K$^{-2}$\\
$\sigma^{2}_{A_{2}}$ & \SI{5.91e6}{} & eV$^{-4}$cm$^{-2}$\\
$\sigma^{2}_{E_{g1}(0)}$ & \SI{8.50e-6}{} & eV$^{2}$\\
\hline
Covariance \hspace{0.7cm} & & \\
\hline
$\sigma_{c_{0}}\sigma_{c_{1}}$ & \SI{2.67e-4}{} & eV$^{-2}$cm$^{-1}$K$^{-1}$\\
$\sigma_{c_{0}}\sigma_{A_{2}}$ & \SI{-1.73e3}{} & eV$^{-4}$cm$^{-2}$\\
$\sigma_{c_{0}}\sigma_{E_{g1}(0)}$ & \SI{1.25e-2}{} & eV$^{-1}$cm$^{-1}$\\
$\sigma_{c_{1}}\sigma_{A_{2}}$ & \SI{-1.23e-2}{} & eV$^{-2}$cm$^{-1}$K$^{-1}$\\
$\sigma_{c_{1}}\sigma_{E_{g1}(0)}$ & \SI{-2.97e-8}{} & eV$^{-1}$K$^{-1}$\\
$\sigma_{A_{2}}\sigma_{E_{g1}(0)}$ & \SI{-0.426}{} & eV$^{-1}$cm$^{-1}$\\
\hline
\end{tabular}

\label{tab:Egap}
\index{tables}
\end{table}

\section{Ionization Model for Deposition Energies above 50\,eV} \label{ion_app}

For energy depositions ($E_\mathrm{r}$) above 50\,eV, the \eh-pair generation probabilities were calculated using Eqs.~13, 14, and 15 from Ref.~\cite{Ramanathan2020}. For Eqs.~13 and 14, additional precision was provided by the paper authors. The resulting versions used for this analysis were Eqs.~\ref{eq:RK_13} and \ref{eq:RK_14}, respectively. $\epsilon_{\text{eh},\infty}$ is the mean energy per electron-hole-pair created for $E_\mathrm{r}$ above 50\,eV, $F_{\infty}$ is the unitless Fano factor for $E_\mathrm{r}$ above 50\,eV, and and $A = 5.2$ is a phenomenological constant.

\begin{equation} \label{eq:RK_13}
    \epsilon_{\text{eh},\infty} (\text{eV}) = 1.6989 E_\text{gap} (\text{eV}) + 0.0843 A + 1.2972
\end{equation}

\begin{equation} \label{eq:RK_14}
    F_{\infty} = -0.0281 E_\text{gap} (\text{eV}) + 0.0015 A + 0.1383
\end{equation}



\bibliographystyle{apsrev4-2}
\bibliography{refs}

\end{document}

%% file: SuperCDMS_AuthorList.tex
\author{M.F.~Albakry} \affiliation{Department of Physics \& Astronomy, University of British Columbia, Vancouver, BC V6T 1Z1, Canada}\affiliation{TRIUMF, Vancouver, BC V6T 2A3, Canada}
\author{I.~Alkhatib} \affiliation{Department of Physics, University of Toronto, Toronto, ON M5S 1A7, Canada}
\author{D.~Alonso-González} \affiliation{Departamento de F\'{\i}sica Te\'orica, Universidad Aut\'onoma de Madrid, 28049 Madrid, Spain}\affiliation{Instituto de F\'{\i}sica Te\'orica UAM-CSIC, Campus de Cantoblanco, 28049 Madrid, Spain}
\author{D.W.P.~Amaral} \affiliation{Department of Physics, Durham University, Durham DH1 3LE, UK}
\author{J.~Anczarski} \affiliation{SLAC National Accelerator Laboratory/Kavli Institute for Particle Astrophysics and Cosmology, Menlo Park, CA 94025, USA}
\author{T.~Aralis} \email{taralis@slac.stanford.edu} \affiliation{SLAC National Accelerator Laboratory/Kavli Institute for Particle Astrophysics and Cosmology, Menlo Park, CA 94025, USA}
\author{T.~Aramaki} \affiliation{Department of Physics, Northeastern University, 360 Huntington Avenue, Boston, MA 02115, USA}
\author{I.J.~Arnquist} \affiliation{Pacific Northwest National Laboratory, Richland, WA 99352, USA}
\author{I.~Ataee~Langroudy} \affiliation{Department of Physics and Astronomy, and the Mitchell Institute for Fundamental Physics and Astronomy, Texas A\&M University, College Station, TX 77843, USA}
\author{E.~Azadbakht} \affiliation{Department of Physics and Astronomy, and the Mitchell Institute for Fundamental Physics and Astronomy, Texas A\&M University, College Station, TX 77843, USA}
% ...
